\documentclass[graybox]{svmult}

\usepackage{mathptmx}       
\usepackage{helvet}         
\usepackage{courier}        
\usepackage{type1cm}        
\usepackage{makeidx}         
\usepackage{graphicx}        
\usepackage{multicol}        
\usepackage[bottom]{footmisc}

\makeindex             

\usepackage{amsmath,amsfonts}
\usepackage{listings}
\usepackage{url}
\usepackage[dvips]{epsfig,psfrag}
\usepackage{booktabs}
\usepackage{paralist}
\usepackage[round]{natbib}\renewcommand{\cite}[1]{\citep{#1}} 
\newcommand{\reffig}[1]{{Fig.~\ref{#1}}}
\newcommand{\brefsec}[1]{{Section~\ref{#1}}}

\newcommand{\refeqn}[1]{{(\ref{#1})}}

\newcommand{\PaperXXlangLong}{Forward And Reverse Fortran Extension Language}
\newcommand{\PaperXXlang}{\textsc{Farfel}}
\newcommand{\PaperXXpreprocessor}{\textsc{Farfallen}}
\newcommand{\PaperXXfort}{\textsc{Fortran}}
\newcommand{\PaperXXalgol}{\textsc{Algol}}
\newcommand{\PaperXXscheme}{\textsc{Scheme}}
\newcommand{\PaperXXadifor}{\textsc{Adifor}}
\newcommand{\PaperXXtapenade}{\textsc{Tapenade}}

\newcommand{\PaperXXVLAD}{\textsc{vlad}}
\newcommand{\PaperXXStalingrad}{\textsc{Stalingrad}}
\newcommand{\PaperXXsyntax}[1]{\textrm{\emph{\normalsize\underline{#1}}}}
\newcommand{\PaperXXsyntaxx}[3]{\textrm{\emph{\normalsize\underline{#1}#2\underline{#3}}}}
\newcommand{\PaperXXvar}{\PaperXXsyntax{var}}
\newcommand{\PaperXXexpr}{\PaperXXsyntaxx{ex}{p}{r}}
\newcommand{\PaperXXstatements}{\PaperXXsyntax{statements}}
\newcommand{\PaperXXTANGENT}[1]{\acute{#1}}
\newcommand{\PaperXXCOTANGENT}[1]{\grave{#1}}
\newcommand{\PaperXXpar}[1]{%
\mbox{}
\par
\noindent
\textbf{#1} \\
}

\DeclareMathOperator*{\argmax}{argmax}

\definecolor{darkblue}{rgb}{0,0,0.7}
\definecolor{darkgreen}{rgb}{0,0.5,0}
\definecolor{darkred}{rgb}{0.7,0,0}

\newcommand{\PaperXXFORWARDJ}{{\color{darkgreen}\overrightarrow{\mathbf{J}}}}
\newcommand{\PaperXXREVERSEJ}{{\color{darkred}\overleftarrow{\mathbf{J}}}}

\begin{document}

{
\title*{AD \emph{in} Fortran}
\subtitle{Part 2: Implementation \emph{via} Prepreprocessor}
\titlerunning{AD \emph{in} Fortran: Implementation}
\author{Alexey Radul \and Barak A. Pearlmutter \and Jeffrey Mark Siskind}
\institute{
Alexey Radul \at Hamilton Inst., National Univ.\ Ireland
Maynooth,
\url{alexey.radul@nuim.ie}
\and
Barak A. Pearlmutter \at Dept.\ Computer Sci.\ \& Hamilton Inst., National Univ.\ Ireland
Maynooth,
\url{barak@cs.nuim.ie}
\and
Jeffrey Mark Siskind \at Electrical and Computer Engineering, Purdue
University, IN, USA, \url{qobi@purdue.edu}}
\maketitle

\abstract{We describe an implementation of the
  \PaperXXlang\ \PaperXXfort\ AD
  extensions \cite{PaperXXdesign}.
  These extensions integrate forward and reverse AD
  directly into the programming model, with attendant benefits to
  flexibility, modularity, and ease of use.   The
  implementation
  we describe
  is a ``prepreprocessor'' that
  generates input to
  existing \PaperXXfort-based AD tools.
  In essence, blocks of code which
  are targeted for AD by \PaperXXlang\ constructs
  are put into subprograms which capture their lexical
  variable context, and these are closure-converted into top-level
  subprograms and specialized to eliminate \lstinline{EXTERNAL} arguments,
  rendering them amenable to existing AD preprocessors, which are then
  invoked, possibly repeatedly if the AD is nested.
}

\section{Introduction}\label{PaperXXsec:intro}

The \emph{\PaperXXlangLong} (\PaperXXlang)
extensions to \PaperXXfort\ enable smooth and modular use of AD  \cite{PaperXXdesign}.
A variety of implementation strategies present themselves,
ranging from
\begin{inparaenum}[(a)]
\item deep integration into a \PaperXXfort\ compiler, to
\item a preprocessor that performs the requested AD and generates
\PaperXXfort\ or some other high-level language, to
\item a prepreprocessor which itself does no AD but generates input to an existing
\PaperXXfort\ AD preprocessor.
\end{inparaenum}
This last strategy
leverages existing \PaperXXfort-based AD tools and
compilers,
avoiding re-implementation of the AD
transformations,
at the expense of inheriting some of the limitations of
the AD tool it invokes.
This is the technique we will focus upon.

\PaperXXpreprocessor\ transforms \PaperXXlang\ input
into \PaperXXfort, and invokes an existing
AD system \cite{Bischof1992AGD, Hascoet2004TUG} to generate the needed
derivatives.
The process can make use of a variety of existing
\PaperXXfort-based AD preprocessors, making it easy for the
programmer to switch between them.
There is significant semantic mismatch between \PaperXXlang\ and the
AD operations allowed by the AD systems used, necessitating rather
dramatic code transformations.
When the \PaperXXlang\ program involves nested AD, the
transformations and staging become even more involved.
Viewed as a whole, this tool automates the task of applying AD,
including the
detailed maneuvers required for nested application of existing tools,
thereby extending the reach and
utility of AD.\@

The remainder of the paper is organized as follows:
\brefsec{PaperXXsec:language} reviews the \PaperXXlang\ extensions.
A complete example program on
page~\pageref{PaperXXexample} illustrates their use.
\brefsec{PaperXXsec:implementation} describes the implementation in
detail using this example program.
\brefsec{PaperXXsec:conclusion} summarizes  work's contributions.

\section{Language Extensions}\label{PaperXXsec:language}

\PaperXXlang\ provides two principal
extensions to \PaperXXfort: syntax for AD and for nested subprograms.

\PaperXXpar{Extension 1: AD Syntax}
\PaperXXlang\ adds the \lstinline{ADF} construct for forward
AD:
\begin{lstlisting}
      ADF(TANGENT($\PaperXXvar$) = $\PaperXXexpr$ $\ldots$)
      $\PaperXXstatements$
      END ADF($\PaperXXvar$ = TANGENT($\PaperXXvar$) $\ldots$)
\end{lstlisting}
Multiple opening and closing assignments are separated by commas.
Independent variables are listed in the ``calls'' to \lstinline{TANGENT} on
the left-hand sides of the opening assignments and are given the specified
tangent values.
Dependent variables appear in the ``calls'' to \lstinline{TANGENT} on the
right-hand sides of the closing assignments and the corresponding tangent
values are assigned to the indicated destination variables.
The \lstinline{ADF} construct uses forward AD to compute the directional
derivative of the dependent variables at the point specified by the vector of
independent variables in the direction specified by the vector of tangent
values for the independent variables and assigns it to the destination
variables.

An analogous \PaperXXlang\  construct supports reverse AD:
\begin{lstlisting}
      ADR(COTANGENT($\PaperXXvar$) = $\PaperXXexpr$ $\ldots$)
      $\PaperXXstatements$
      END ADR($\PaperXXvar$ = COTANGENT($\PaperXXvar$) $\ldots$)
\end{lstlisting}
\emph{Dependent} variables are listed in the ``calls'' to
\lstinline{ COTANGENT} on left-hand sides of the opening assignments and are
given the specified cotangent values as inputs to the reverse phase.
\emph{Independent} variables appear in the ``calls'' to
\lstinline{ COTANGENT} on the right-hand sides of the closing assignments and
the corresponding cotangent values at the end of the reverse phase are
assigned to the indicated destination variables.
The \lstinline{ADR} construct uses reverse AD to compute the
gradient with respect to the independent variables at the point specified by
the vector of
independent variables induced by the specified gradient with respect
to the dependent variables, and assigns it to the destination variables.
The expressions used to initialize the cotangent inputs to the reverse
phase are evaluated at the end of the forward phase, even though they
appear textually prior to the statements specifying the forward phase.
This way, the direction input to the reverse phase can depend on the
result of the forward phase.

For both \lstinline{ADF} and \lstinline{ADR}, implied-\lstinline{DO} syntax is
used to allow arrays in the opening and closing assignments.
By special dispensation, the statement \lstinline{ADF($\PaperXXvar$)}
is interpreted as \lstinline{ADF(TANGENT($\PaperXXvar$)=1)} and
\lstinline{ADR($\PaperXXvar$)} as
\lstinline{ADR(COTANGENT($\PaperXXvar$)=1)}.

\PaperXXpar{Extension 2: Nested Subprograms}
In order to conveniently support distinctions between different
variables of differentiation for distinct invocations of AD, as in the
example below,
we borrow from \PaperXXalgol~60 \cite{PaperXXNaur-1963xxx} and generalize
the \PaperXXfort\ ``statement function'' construct by allowing
subprograms to be defined inside other
subprograms, with lexical scope.
As in \PaperXXalgol~60, the scope of parameters and declared variables
is the local subprogram, and these may shadow identifiers from the
surrounding scope.
Implicitly declared variables have the top-level subprogram as their scope.

\PaperXXpar{Concrete Example}
In order to describe the implementation of the above
constructs, we employ a concrete example.
The task is to find an equilibrium $(a^*, b^*)$ of a
two-player game with continuous scalar strategies $a$ and $b$ and
given payoff functions $A$ and $B$.  The method is to find roots of
\begin{equation}\label{PaperXXeqn:game}
  a^* = \argmax_a A(a,\argmax_b B(a^*,b))
\end{equation}
The full program is given, for reference,
in Listing~\ref{PaperXXexample}.
The heart of the program is the implementation \lstinline{EQLBRM}
of \refeqn{PaperXXeqn:game}.
\begin{lstlisting}[float,caption={Complete example \PaperXXlang\ program: equilibria of a continuous-strategy game.},label=PaperXXexample]
C     ASTAR & BSTAR: GUESSES IN, OPTIMIZED VALUES OUT
      SUBROUTINE EQLBRM(BIGA, BIGB, ASTAR, BSTAR, N)
      EXTERNAL BIGA, BIGB
        FUNCTION F(ASTAR)
          FUNCTION G(A)
            FUNCTION H(B)
            H = BIGB(ASTAR, B)
            END
          BSTAR = ARGMAX(H, BSTAR, N)
          G = BIGA(A, BSTAR)
          END
        F = ARGMAX(G, ASTAR, N)-ASTAR
        END
      ASTAR = ROOT(F, ASTAR, N)
      END

      FUNCTION ROOT(F, X0, N)
      X = X0
      DO 1669 I=1,N
      CALL DERIV2(F, X, Y, YPRIME)
 1669 X = X-Y/YPRIME
      ROOT = X
      END

      SUBROUTINE DERIV2(F, X, Y, YPRIME)
      EXTERNAL F
      ADF(X)
      Y = F(X)
      END ADF(YPRIME = TANGENT(Y))
      END

      FUNCTION ARGMAX(F, X0, N)
        FUNCTION FPRIME(X)
        FPRIME = DERIV1(F, X)
        END
      ARGMAX = ROOT(FPRIME, X0, N)
      END

      FUNCTION DERIV1(F, X)
      EXTERNAL F
      ADF(X)
      Y = F(X)
      END ADF(DERIV1 = TANGENT(Y))
      END

      FUNCTION GMBIGA(A, B)
      PRICE = 20-0.1*A-0.1*B
      COSTS = A*(10-0.05*A)
      GMBIGA = A*PRICE-COSTS
      END

      FUNCTION GMBIGB(A, B)
      PRICE = 20-0.1*B-0.0999*A
      COSTS = B*(10.005-0.05*B)
      GMBIGB = B*PRICE-COSTS
      END

      PROGRAM MAIN
      READ *, ASTAR
      READ *, BSTAR
      READ *, N
      CALL EQLBRM(GMBIGA, GMBIGB, ASTAR, BSTAR, N)
      PRINT *, ASTAR, BSTAR
      END
\end{lstlisting}
Note that this whole program is only 63 lines of code, with
plenty of modularity boundaries.
This code is used as a running example for the remainder of the paper.

\section{Implementation}\label{PaperXXsec:implementation}

\PaperXXlang\ is implemented by
the \PaperXXpreprocessor\ preprocessor.
The current version
is merely a proof of concept, and not production quality: it does
not accept the entire \PaperXXfort{}77 language, and does not scale.  However,
its principles of operation will be unchanged in a forthcoming
production-quality implementation.  Here we describe the reduction
of \PaperXXlang\ constructs to \PaperXXfort, relying on existing
\PaperXXfort-based
AD tools for the actual derivative transformations.

\PaperXXlang\ introduces two new constructs into \PaperXXfort:
nested subprograms and syntax for requesting AD.\@  We
implement nested subprograms by extracting them to the top level, and
communicating the free variables from the enclosing
subprogram by passing them as arguments into the new top-level
subprogram.  This is an instance of \emph{closure conversion},
a standard class of techniques
for converting nested subprograms to top-level ones \cite{johnsson85lambda}.
In order
to accommodate passing formerly-free variables as arguments,
we must adjust all the call sites of the
formerly-nested subprogram; we must specialize all the subprograms that
accept that subprogram as an external to also accept the extra closure
parameters; and adjust all call sites to all those specialized subprograms to
pass those extra parameters.

We implement the AD syntax by constructing new subroutines that
correspond to the statements inside each \lstinline{ADF} or
\lstinline{ADR} block, arranging to call the AD tool of choice on each
of those new subroutines, and transforming the block itself into a
call to the appropriate tool-generated subroutine.

\PaperXXpar{Nested Subprograms in Detail}
Let us illustrate closure conversion on our example.
Recall \lstinline{ARGMAX} in our example program:
\begin{lstlisting}
      FUNCTION ARGMAX(F, X0, N)
        FUNCTION FPRIME(X)
        FPRIME = DERIV1(F, X)
        END
      ARGMAX = ROOT(FPRIME, X0, N)
      END
\end{lstlisting}
This contains the nested function \lstinline{FPRIME}.
We closure convert this as follows.
First, extract \lstinline{FPRIME}
to the top level:
\begin{lstlisting}
      FUNCTION ARGMAX_FPRIME(X, F)
      ARGMAX_FPRIME = DERIV1(F, X)
      END

      FUNCTION ARGMAX(F, X0, N)
      ARGMAX = ROOT_1(ARGMAX_FPRIME, F, X0, N)
      END
\end{lstlisting}
Note the addition of a closure argument for \lstinline{F} since it is freely
referenced in \lstinline{FPRIME}, and the addition of the same closure
argument at the call site, since
\lstinline{FPRIME} is passed as an external to \lstinline{ROOT}.
Then we specialize \lstinline{ROOT} to create a version
that accepts the needed set of closure arguments (in this case
one):
\begin{lstlisting}
      FUNCTION ROOT_1(F, F1, X0, N)
      X = X0
      DO 1669 I=1,N
      CALL DERIV2_1(F, F1, X, Y, YPRIME)
 1669 X = X-Y/YPRIME
      ROOT_1 = X
      END
\end{lstlisting}
Since \lstinline{ROOT} contained a call to \lstinline{DERIV2}, passing
it the external passed to \lstinline{ROOT}, we must also specialize
\lstinline{DERIV2}:
\begin{lstlisting}
      SUBROUTINE DERIV2_1(F, F1, X, Y, YPRIME)
      EXTERNAL F
      ADF(X)
      Y = F(X, F1)
      END ADF(YPRIME = TANGENT(Y))
      END
\end{lstlisting}
We must, in general, copy and specialize the portion of the call graph
where the nested subprogram travels, which in this case is just two
subprograms.
During such copying and specialization, we propagate external
constants (e.g., \lstinline{FPRIME} through the call to \lstinline{ROOT_1},
the call to \lstinline{DERIV2_1}, and the call site therein) allowing the
elimination of the \lstinline{EXTERNAL} declaration for these values.
This supports AD tools that do not allow taking derivatives through calls to
external subprograms.



That is the process for handling one nested subprogram.  In our
example, the same is done for \lstinline{F} in
\lstinline{EQLBRM}, \lstinline{G} in \lstinline{F}, and \lstinline{H}
in \lstinline{G}.
Doing so causes the introduction of a number of closure arguments, and
the specialization of a number of subprograms to accept those arguments;
including perhaps further specializing things that have already been
specialized.
The copying also allows a limited form of subprogram reentrancy:
even if recursion is disallowed (as in traditional \PaperXXfort{}77)
our nested uses of \lstinline{ARGMAX} will cause no difficulties
because they will end up calling two different specializations of
\lstinline{ARGMAX}.

Note that we must take care to prevent this process from introducing
spurious aliases.
For example, in \lstinline{EQLBRM}, the internal
function \lstinline{F} that is passed to \lstinline{ROOT} closes over
the iteration count \lstinline{N}, which is also passed to
\lstinline{ROOT} separately.
When specializing \lstinline{ROOT} to accept the closure parameters
of \lstinline{F}, we must not pass \lstinline{N} to the specialization
of \lstinline{ROOT} twice, lest we run afoul of \PaperXXfort's prohibition
against assigning to aliased values.
Fortunately, such situations are syntactically apparent.

Finally, specialization leaves behind unspecialized (or
underspecialized) versions of subprograms, which may now be unused,
and if so can be eliminated.
In this case, that includes \lstinline{ROOT}, \lstinline{DERIV2},
\lstinline{ARGMAX}, \lstinline{FPRIME}, and \lstinline{DERIV1} from
the original program, as well as some intermediate specializations
thereof.

\PaperXXpar{AD Syntax in Detail}
We implement the AD syntax by first canonicalizing each
\lstinline{ADF} or \lstinline{ADR} block to be a single call to a
(new, internal) subroutine, then extracting those subroutines to
the top level, then rewriting the block to be a call to an
AD-transformed version of the subroutine, and then arranging to call the AD
tool of choice on each of those new subroutines to generate the needed
derivatives.

Returning to our example program, closure conversion of nested
subprograms produced the following specialization of
\lstinline{DERIV1}:
\begin{lstlisting}
      FUNCTION DERIV1_1(ASTAR, BIGA, BIGB, BSTAR, N, X)
      ADF(X)
      Y = EQLBRM_F_G(X, ASTAR, BIGA, BIGB, BSTAR, N)
      END ADF(DERIV1_1 = TANGENT(Y))
      END
\end{lstlisting}
which contains an \lstinline{ADF} block.  We seek to convert this
into a form suitable for invoking the AD preprocessor.  We first canonicalize
by introducing a new subroutine to capture the statements in
the \lstinline{ADF} block, producing the following:
\begin{lstlisting}
      FUNCTION DERIV1_1(ASTAR, BIGA, BIGB, BSTAR, N, X)
        SUBROUTINE ADF1()
        Y = EQLBRM_F_G(X, ASTAR, BIGA, BIGB, BSTAR, N)
        END
      ADF(X)
      CALL ADF1()
      END ADF(DERIV1_1 = TANGENT(Y))
      END
\end{lstlisting}
Extracting the subroutine \lstinline{ADF1} to the top level as before yields
the following:
\begin{lstlisting}
      SUBROUTINE DERIV1_1_ADF1(X, ASTAR, BIGA, BIGB, BSTAR, N, Y)
      Y = G(X, ASTAR, BIGA, BIGB, BSTAR, N)
      END

      FUNCTION DERIV1_1(ASTAR, BIGA, BIGB, BSTAR, N, X)
      ADF(X)
      CALL DERIV1_1_ADF1(X, ASTAR, BIGA, BIGB, BSTAR, N, Y)
      END ADF(DERIV1_1 = TANGENT(Y))
      END
\end{lstlisting}
Now we are properly set up to rewrite the \lstinline{ADF} block into
a subroutine call---specifically, to a subroutine that will be generated
from \lstinline{DERIV1_1_ADF1} by AD.\@
The exact result depends on the AD tool that will be used to construct
the derivative of \lstinline{DERIV1_1_ADF1}; for \PaperXXtapenade, the
generated code looks like this:
\begin{lstlisting}
      FUNCTION DERIV1_1(ASTAR, BIGA, BIGB, BSTAR, N, X)
      X_G1 = 1
      ASTAR_G1 = 0
      BSTAR_G1 = 0
      CALL DERIV1_1_ADF1_G1(X, X_G1, ASTAR, ASTAR_G1, BIGA, BIGB,
     +BSTAR, BSTAR_G1, N, Y, DERIV1_1)
      END
\end{lstlisting}
Different naming conventions are used for \lstinline{DERIV1_1_ADF1_G1} when
generating
code for \PaperXXadifor; the parameter passing conventions of \PaperXXtapenade\
and \PaperXXadifor\ agree in this case.
\PaperXXpreprocessor\ maintains the types of variables in order to know whether
to generate variables to hold tangents and cotangents (which are
initialized to zero if they were not declared in the relevant opening
assignments.)

The same must be repeated for each \lstinline{ADF} and \lstinline{ADR}
block; in our example there are five in all: two in specializations
of \lstinline{DERIV1} and three in specializations of \lstinline{DERIV2}.
We must also specialize \lstinline{EQLBRM} and its descendants in the
call graph, by the process already illustrated, to remove
external calls to the objective functions, for the reasons
described earlier.

Finally, we must invoke the user's preferred AD tool to generate
all the needed derivatives.
Here, \PaperXXpreprocessor\ might invoke \PaperXXtapenade\ as follows:
\begin{lstlisting}[language=bash,frame=trbl,frameround=,aboveskip=1ex,belowskip=1ex]
#! /bin/sh
tapenade -root deriv1_2_adf2 -d -o eqlbrm42 -diffvarname "_g2"\
    -difffuncname "_g2" eqlbrm42.f
tapenade -root deriv2_1_2_adf4 -d -o eqlbrm42 -diffvarname "_g4"\
    -difffuncname "_g4" eqlbrm42{,_g2}.f
tapenade -root deriv1_1_adf1 -d -o eqlbrm42 -diffvarname "_g1"\
    -difffuncname "_g1" eqlbrm42{,_g2,_g4}.f
tapenade -root deriv2_1_1_adf3 -d -o eqlbrm42 -diffvarname "_g3"\
    -difffuncname "_g3" eqlbrm42{,_g2,_g4,_g1}.f
tapenade -root deriv2_2_adf5 -d -o eqlbrm42 -diffvarname "_g5"\
    -difffuncname "_g5" eqlbrm42{,_g2,_g4,_g1,_g3}.f
\end{lstlisting}
We must take care that multiple invocations of
the AD tool to generate the various derivatives occur in the proper
order, which is computable from the call graph
of the program, to ensure that generated derivative codes that are used
in subprograms to be differentiated are available to be transformed.
For example, all the derivatives of \lstinline{EQLBRM_F_G_H} needed
for its optimizations must be generated before generating derivatives
of (generated subprograms that call) \lstinline{EQLBRM_F_G}.
We must also take care to invoke the AD tool with different prefixes/suffixes,
so that variables and subprograms created by one differentiation do
not clash with those made by another.

\PaperXXpar{Performance}
We tested the performance of code generated by \PaperXXpreprocessor\ in
conjunction with \PaperXXtapenade\ and with \PaperXXadifor.  We executed
\PaperXXpreprocessor\ once to generate \PaperXXfort{}77 source using
\PaperXXtapenade\ for the automatic differentiation, and separately
using \PaperXXadifor\ for the AD.\@  In each case, we compiled the
resulting program
(\texttt{gfortran} 4.6.2-9, 64-bit Debian
sid, \texttt{-Ofast} \texttt{-fwhole-program}, single precision)
with $N=1000$ iterations at each level
and timed the execution
of the binary
on a 2.93GHz Intel i7 870.
For comparison, we translated the same computation into \PaperXXVLAD\
\cite{Pearlmutter2008UPL} (\reffig{PaperXXfig:vlad}), compiled it with
\PaperXXStalingrad~\cite{Siskind2008UPU}, and ran it on the same machine.
\PaperXXStalingrad\ has the
perhaps unfair advantage of being an optimizing compiler with
integrated support for AD, so we are pleased that \PaperXXpreprocessor\
was able to achieve performance that was nearly competitive.
\begin{center}
\begin{tabular}{c@{\qquad}c@{\qquad}c}\toprule
\PaperXXtapenade & \PaperXXadifor & \PaperXXStalingrad \\\midrule
 6.97 & 8.92 & 5.83 \\\bottomrule
\end{tabular}
\end{center}

\begin{figure}[tb]
\newcommand{\define}{\stackrel{{\color{darkblue}\triangle}}{{\color{darkblue}=}}}
\newcommand{\IF}{\textbf{if}\;}
\newcommand{\THEN}{\textbf{then}\;}
\newcommand{\ELSE}{\textbf{else}\;}
\newcommand{\LET}{\textbf{let}\;}
\newcommand{\IN}{\textbf{in}\;}
\centering
\begin{equation*}
\begin{array}{@{}l@{}}
\textsc{eqlbrm}(A,B,a_0,b_0,n)\define
\begin{array}[t]{@{}l@{}}
\LET\begin{array}[t]{@{}l@{}}
     f(a')\define\begin{array}[t]{@{}l@{}}
                 \LET g(a)\define\begin{array}[t]{@{}l@{}}
                                 \LET h(b)\define B(a',b)\\
                                 \IN A(a,\textsc{argmax}(h,b_0,n))
                                 \end{array}\\
                 \IN\textsc{argmax}(g,a',n)-a'
                 \end{array}\\
     a^{*}=\textsc{root}(f,a_0,n)
    \end{array}\\
\IN\begin{array}[t]{@{}l@{}}
    \LET\begin{array}[t]{@{}l@{}}
         h(b)\define B(a^{*},b)\\
         b^{*}=\textsc{argmax}(h,b_0,n)
        \end{array}\\
    \IN a^{*},b^{*}
    \end{array}
\end{array}\\
\textsc{root}(f,x,n)\define\begin{array}[t]{@{}l@{}}
                           \IF n=0\\
                           \THEN x\\
                           \ELSE\begin{array}[t]{@{}l@{}}
                                \LET y,y'=\textsc{deriv2}(f,x)\\
                                \IN\textsc{root}(f,x-\frac{y}{y'},n-1)
                                \end{array}
                           \end{array}\\
\textsc{deriv2}(f,x)\define\PaperXXFORWARDJ(f,x,1)\\
\textsc{argmax}(f,x_0,n)\define\begin{array}[t]{@{}l@{}}
                               \LET f'(x)\define\textsc{deriv1}(f,x)\\
                               \IN\textsc{root}(f',x_0,n)
                               \end{array}\\
\textsc{deriv1}(f,x)\define\begin{array}[t]{@{}l@{}}
                           \LET x,\acute{y}=\PaperXXFORWARDJ(f,x,1)\\
                           \IN\acute{y}
                           \end{array}\\
\textsc{a}(a,b)\define\begin{array}[t]{@{}l@{}}
                      \LET\begin{array}[t]{@{}l@{}}
                          \textit{price}=20-0.1\times a-0.1\times b\\
                          \textit{costs}=a\times(10-0.05\times a)
                          \end{array}\\
                      \IN a\times\textit{price}-\textit{costs}\\
                      \end{array}\\
\textsc{b}(a,b)\define\begin{array}[t]{@{}l@{}}
                      \LET\begin{array}[t]{@{}l@{}}
                          \textit{price}=20-0.1\times b-0.0999\times a\\
                          \textit{costs}=b\times(10.005-0.05\times b)
                          \end{array}\\
                      \IN b\times\textit{price}-\textit{costs}\\
                      \end{array}\\
\begin{array}[t]{@{}l@{}}
\LET\begin{array}[t]{@{}l@{}}
    a_0={\color{darkblue}\textsc{readReal}}()\\
    b_0={\color{darkblue}\textsc{readReal}}()\\
    n={\color{darkblue}\textsc{readReal}}()\\
    a^{*},b^{*}=\textsc{eqlbrm}(\textsc{a},\textsc{b},a_0,b_0,n)
    \end{array}\\
\IN\begin{array}[t]{@{}l@{}}
   {\color{darkblue}\textsc{writeReal}}(a^{*})\\
   {\color{darkblue}\textsc{writeReal}}(b^{*})
   \end{array}
\end{array}
\end{array}
\end{equation*}
\caption{Complete \PaperXXVLAD\ program for our concrete example with the same
  organization and functionality as the \PaperXXlang\ program in
  Listing~\protect\ref{PaperXXexample}.}
\label{PaperXXfig:vlad}
\end{figure}

The \PaperXXVLAD\ code in \reffig{PaperXXfig:vlad} was written with the same
organization, variable names, subprogram names, and parameter names and order
as the corresponding \PaperXXlang\ code in Listing~\protect\ref{PaperXXexample}
to help a reader unfamiliar with \PaperXXscheme, the language on which
\PaperXXVLAD\ is based, understand how \PaperXXlang\ and \PaperXXVLAD\ both
represent the same essential notions of nested subprograms and the
AD discipline of requesting derivatives precisely where they are needed.
Because functional-programming languages, like \PaperXXscheme\ and
\PaperXXVLAD, prefer higher-order functions (i.e., operators) over the
block-based style that is prevalent in \PaperXXfort, AD is invoked via
the $\PaperXXFORWARDJ$ and $\PaperXXREVERSEJ$ operators rather than via the
\lstinline{ADF} and \lstinline{ADR} statements.
However, there is a one-to-one correspondence between
\begin{equation*}
\PaperXXFORWARDJ:f\times x\times\PaperXXTANGENT{x}\rightarrow y\times\PaperXXTANGENT{y}
\end{equation*}
an operator that takes a function~$f$ as input, along with a primal
argument~$x$ and tangent argument~$\PaperXXTANGENT{x}$ and returns both a
primal result~$y$ and a tangent result~$\PaperXXTANGENT{y}$, and the following:
\begin{lstlisting}
      ADF(TANGENT($x$) = $\PaperXXTANGENT{x}$)
      $y$ = $f$($x$)
      END ADF($\PaperXXTANGENT{y}$ = TANGENT($y$))
\end{lstlisting}
Similarly, there is a one-to-one correspondence between
\begin{equation*}
\PaperXXREVERSEJ:
f\times x\times\PaperXXCOTANGENT{y}\rightarrow y\times\PaperXXCOTANGENT{x}
\end{equation*}
an operator that takes a function~$f$ as input, along with a primal
argument~$x$ and cotangent~$\PaperXXCOTANGENT{y}$ and returns both a
primal result~$y$ and a cotangent~$\PaperXXCOTANGENT{x}$, and
\pagebreak[2]
\begin{lstlisting}
      ADR(COTANGENT($y$) = $\PaperXXCOTANGENT{y}$)
      $y$ = $f$($x$)
      END ADR($\PaperXXCOTANGENT{x}$ = COTANGENT($x$))
\end{lstlisting}
The strong analogy between how the callee-derives AD discipline is represented
in both \PaperXXlang\ and \PaperXXVLAD\ serves two purposes: it enables the
use of the \PaperXXStalingrad\ compiler technology for compiling
\PaperXXlang\ and facilitates migration from legacy imperative languages
to more modern, and more easily optimized, pure languages.

\section{Conclusion}\label{PaperXXsec:conclusion}

We have illustrated an implementation of the \PaperXXlang\ extensions to \PaperXXfort---nested subprograms and syntax
for AD \cite{PaperXXdesign}.
These extensions enable convenient,
modular programming using a callee-derives paradigm of automatic
differentiation.  Our implementation is a preprocessor that translates \PaperXXlang\
\PaperXXfort\ extensions into input suitable
for an existing AD
tool.  This strategy enables modular, flexible use of AD in the
context of an existing legacy language and tool chain, without
sacrificing the desirable performance characteristics of these tools:
in the concrete example,
only 20\%--50\% slower than a
dedicated AD-enabled compiler, depending on which \PaperXXfort\ AD
system is used.

\begin{acknowledgement}
This work was supported, in part, by Science Foundation Ireland grant
09/IN.1/I2637, National Science Foundation grant CCF-0438806, the Naval Research Laboratory under
Contract Number N00173-10-1-G023, and the Army Research Laboratory
accomplished under Cooperative Agreement Number W911NF-10-2-0060.
Any views, opinions, findings, conclusions, or recommendations contained or
expressed in this document or material are those of the author(s) and do not
necessarily reflect or represent the views or official policies, either
expressed or implied, of SFI, NSF, NRL, the Office of Naval Research, ARL, or
the Irish or U.S. Governments.
The U.S. Government is authorized to reproduce and distribute reprints for
Government purposes, notwithstanding any copyright notation herein.
\end{acknowledgement}

\bibliographystyle{plainnat}
\bibliography{../ad,../Papers/PaperXX/paper}

}

\end{document}